\documentclass[twocolumn,showpacs,preprintnumbers,amsmath,amssymb]{revtex4}

\usepackage{graphicx}% Include figure files
\usepackage{dcolumn}% Align table columns on decimal point
\usepackage{bm}% bold math
\usepackage{tabularx}

\newcommand{\muev} {$\mathrm{\mu eV}$}
\newcommand{\Mpl} {M_{\mathrm{pl}}}
\newcommand{\Pin} {P_{\mathrm{in}}}
\newcommand{\Pout} {P_{\mathrm{out}}}
\newcommand{\TEmode} {$\mathrm{TE_{011}}$}

\begin{document}

\title{A Search for Scalar Chameleons with ADMX}

\author{G. Rybka, M. Hotz, L. J Rosenberg}
\affiliation{University of Washington, Seattle, Washington 98195}
\author{S.J. Asztalos\footnote{Currently at XIA LLC, 31057 Genstar Rd., Hayward CA, 94544.}, G. Carosi, C. Hagmann, D. Kinion and K. van Bibber\footnote{Currently at Naval Postgraduate School, Monterey, CA}}
\affiliation{Lawrence Livermore National Laboratory, Livermore, California, 94550}
\author{J. Hoskins, C. Martin, P. Sikivie and D.B. Tanner}
\affiliation{University of Florida, Gainesville, Florida 32611}
\author{R. Bradley}
\affiliation{National Radio Astronomy Observatory, Charlottesville, Virginia 22903}
\author{J. Clarke}
\affiliation{University of California and Lawrence Berkeley National Laboratory, Berkeley, California 94720}

\date{\today}

\begin{abstract}
Scalar fields with a ``chameleon" property, in which the effective particle mass is a function of its local environment, are common to many theories beyond the standard model and could be responsible for dark energy.
If these fields couple weakly to the photon, they could be detectable through the ``afterglow" effect of photon-chameleon-photon transitions.
The ADMX experiment was used in the first chameleon search with a microwave cavity to set a new limit on scalar chameleon-photon coupling $\beta_\gamma$ excluding values between $2\times10^{9}$ and $5\times10^{14}$ for effective chameleon masses between 1.9510 and 1.9525 $\mu$eV.

\end{abstract}

%07.57.Kp  Detectors, Microwave
%14.70.Bh photons, properties of
%95.36.+x Dark Energy
%12.20.Fv Quantum electrodynamics - 	Experimental tests

\maketitle

%I. Introduction
%	A) What are Chameleons & What are they good for

Astrophysical observations from a variety of sources all suggest that the expansion of the universe is accelerating \cite{astroreview_darkenergy}.  The negative pressure required for this phenomenon, under the name dark energy, can be interpreted as a nonzero cosmological constant, but could also be the signature of a light scalar field slowly rolling down a shallow potential \cite{PhysRevD.37.3406,PhysRevLett.80.1582}.  Light scalar fields are ubiquitous in physics theories beyond the standard model, but have been severely constrained by short-range gravity experiments \cite{adelberger-annrevnucinv}.

It has been suggested, however, that scalar fields with nonlinear self-interactions can have a  ``chameleon" property \cite{PhysRevLett.93.171104} which causes the effective mass of perturbations to the field to be dependent on the local energy density.
This effect can shield all but a thin shell of test masses from the new force carried by a scalar field, significantly relaxing bounds on couplings from gravity experiments while still offering a viable low mass dark energy candidate on cosmological scales \cite{PhysRevD.70.104001,PhysRevD.70.123518,PhysRevD.69.044026}.  A possible effective potential for such a field is \cite{ PhysRevD.70.123518}

\begin{equation}
V_{\mathrm{eff}}(\phi,\vec{x})=
\Lambda^4\exp\left(\frac{\Lambda^n}{\phi^n}\right)
+e^{\frac{\beta\phi}{\Mpl}}\rho_m(\vec{x})
+e^{\frac{\beta_\gamma\phi}{\Mpl}}\rho_\gamma(\vec{x}) .
\end{equation}
Here $\phi$ is the chameleon field, $\beta$ and $\beta_\gamma$ are unitless couplings to matter and photons, $\Mpl$ is the reduced Planck mass ($2.4\times10^{18}$ GeV), $\rho_m$ and $\rho_\gamma$ are the matter and electromagnetic energy densities, and $\Lambda$ and $n$ are model parameters, with $\Lambda\simeq3\times10^{-12}\ \mathrm{GeV}$ for dark energy.  The field $\phi$ minimizes the potential at each location with some value $\phi_0(\vec{x})$, and the mass of excitations of the field is then
\begin{equation}
m_\phi^2(\vec{x})\simeq\frac{\partial^2 }{\partial \phi^2}V(\phi_0(\vec{x}),\vec{x}).
\end{equation}
The experimentally accessible parameters are the coupling strengths $\beta$, $\beta_\gamma$ and the effective mass of the chameleon $m_\phi$ inside the experiment.

\begin{figure*}
\begin{center}
\begin{tabular*}{0.95\textwidth}{@{\extracolsep{\fill}} cccc }
\includegraphics[width=3.8cm]{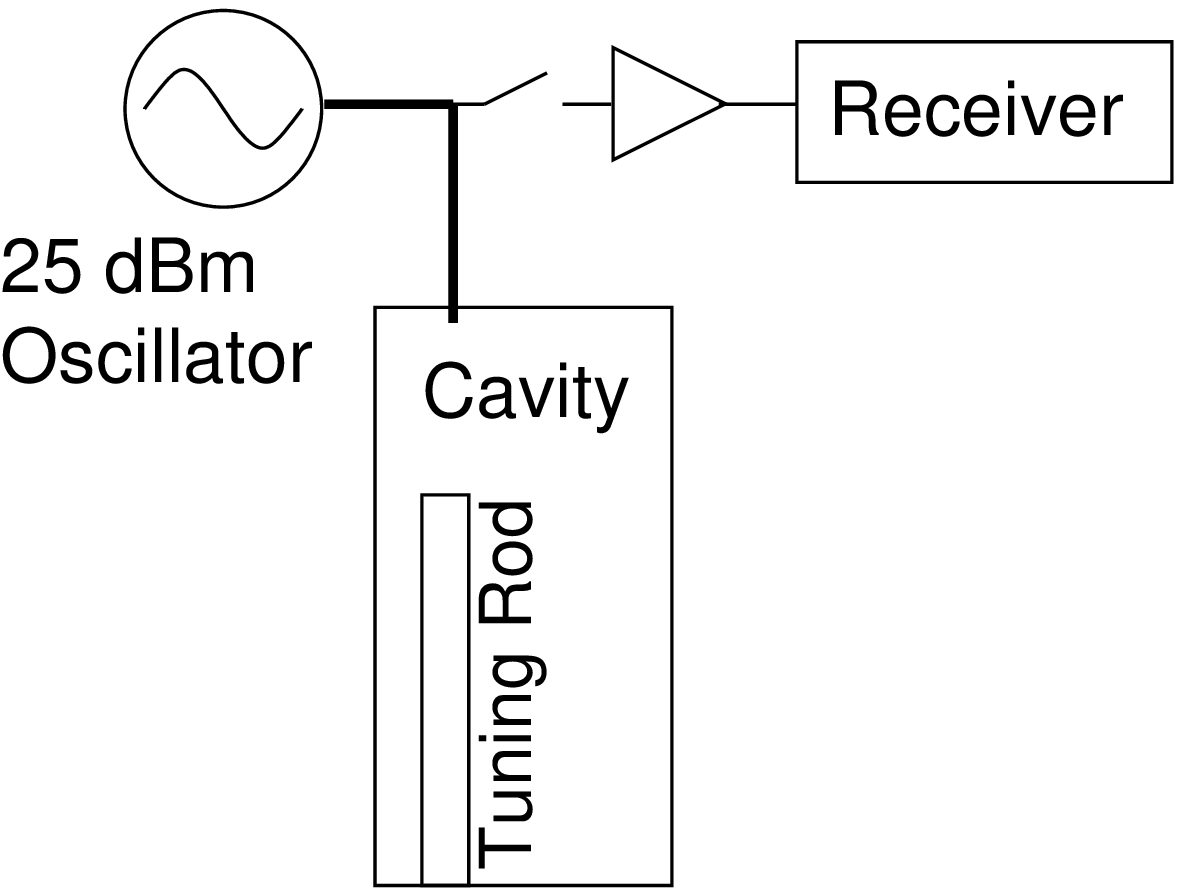} &
\includegraphics[width=3.8cm]{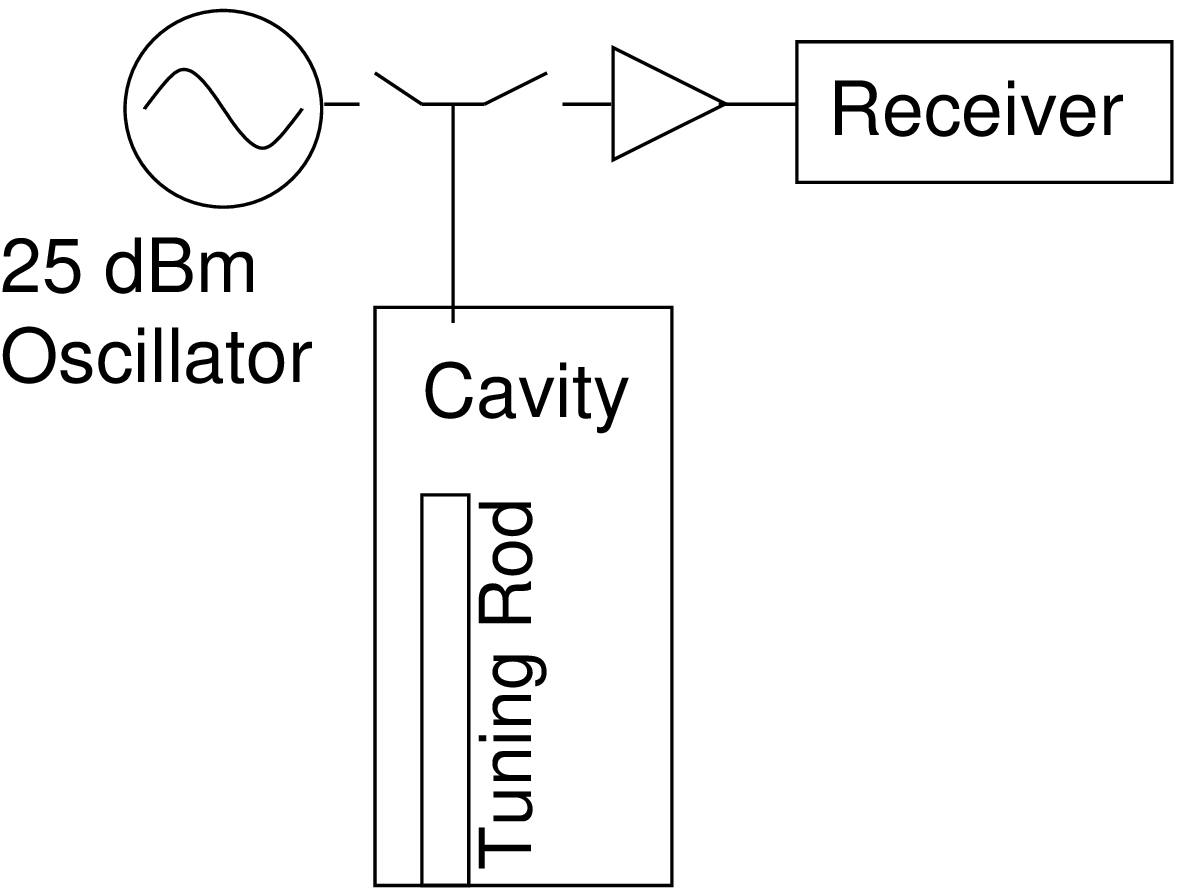} &
\includegraphics[width=3.8cm]{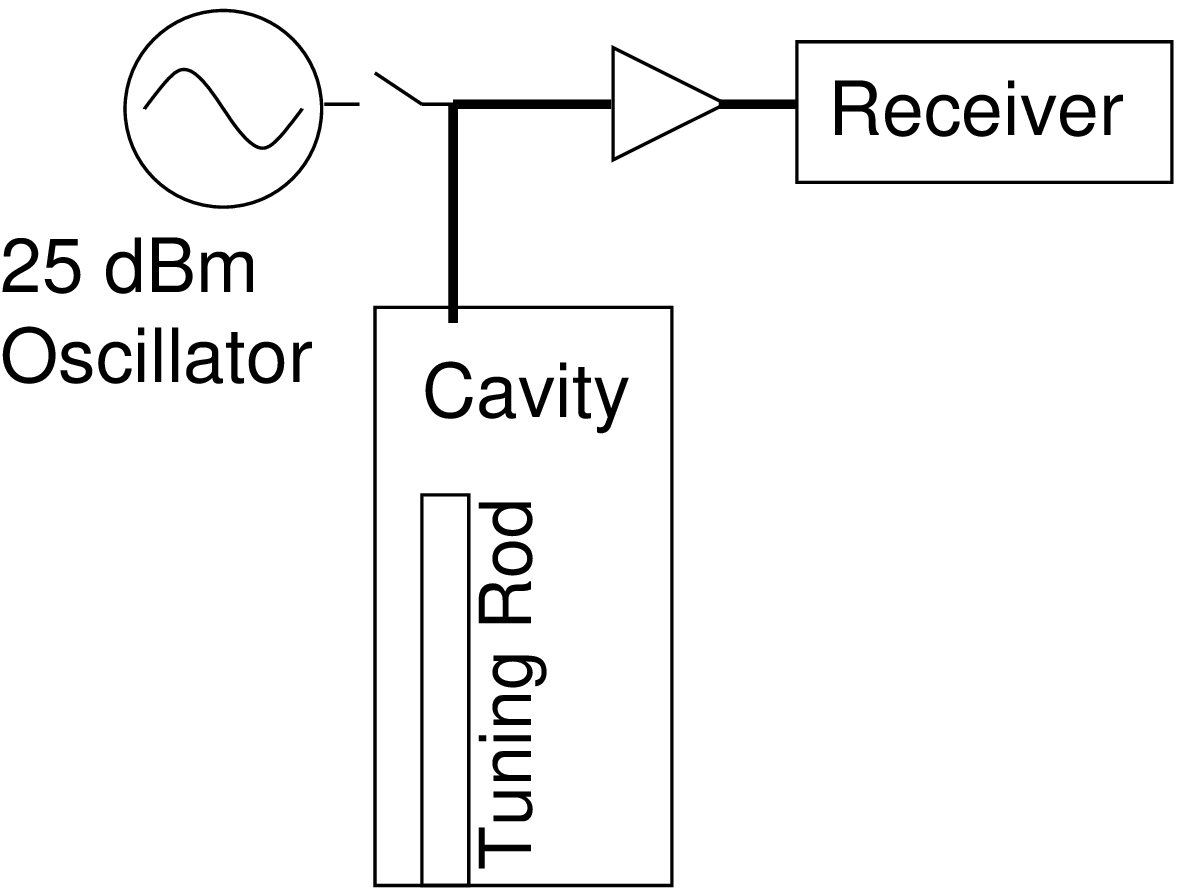} &
\includegraphics[width=3.8cm]{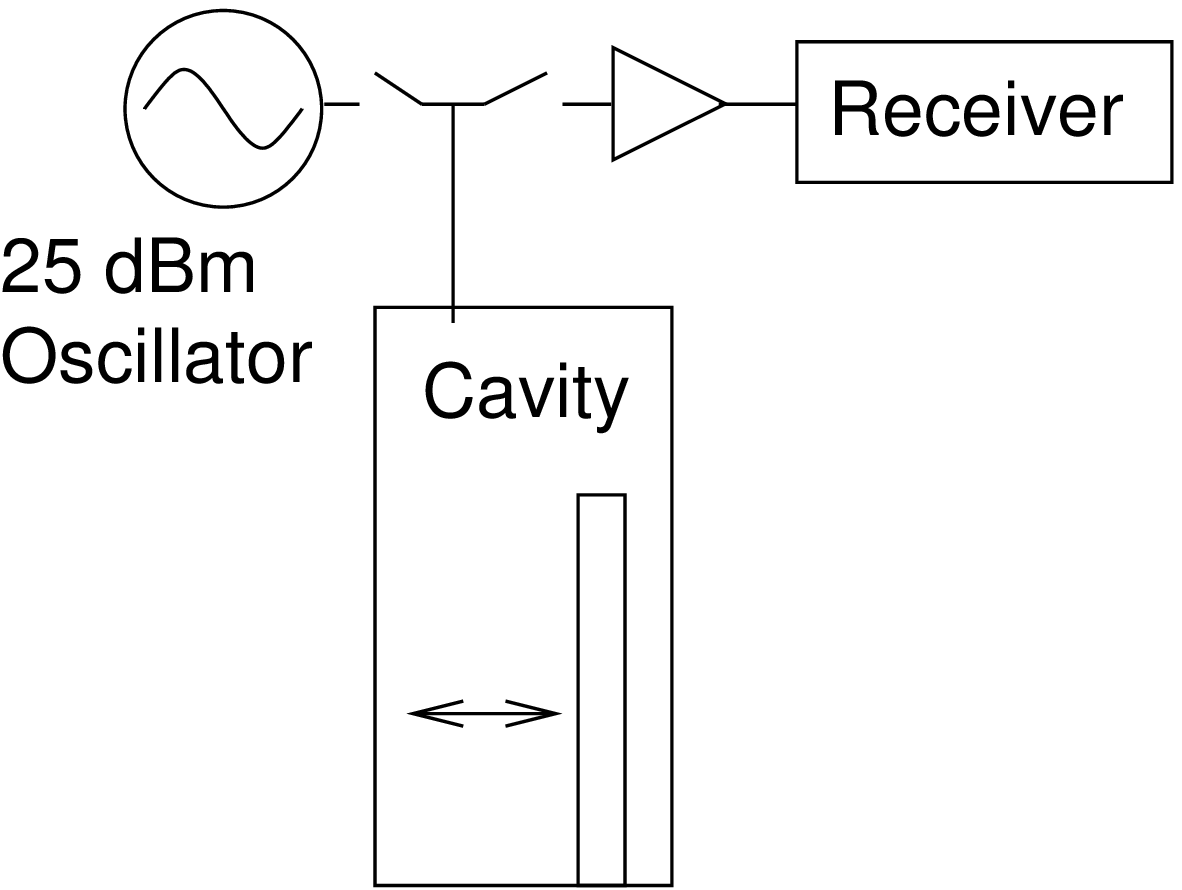}  \\
Step 1 & Step 2 & Step 3 & Step 4 \\
Electromagnetic modes & Pure electromagnetic & Afterglow from chameleon & Tuning rods moved to\\
excited by source & modes decay & modes measured & change frequency \\
\end{tabular*}
\caption{\label{fig:experiment_process} Chameleon search procedure}
\end{center}
\end{figure*}

Scalar chameleons may have a different coupling strength to the electromagnetic field than to matter \cite{PhysRevLett.99.121103}.  If the electromagnetic coupling is dominant, electromagnetic experiments searching for dark energy may be more fruitful than gravitational ones.
For example, laser experiments utilize the unique nature of chameleons to look for the ``afterglow" of photon-chameleon-photon transitions \cite{ahlers:015018,gies:025016}.  In general, these experiments involve shining a laser through a closed, empty container subjected to a large magnetic field.
In the magnetic field, some photons from the laser mix into chameleons.  For models in which the chameleon mass inside the walls of the container is much greater than that in vacuum, the chameleons are trapped because their effective mass in the walls is greater than their total energy.  
If the mixing time between chameleons and photons is longer than the time the photons spend in the container, photons may be detected for a time after the laser is turned off while the trapped chameleons mix back into photons, which subsequently escape the container.
Current limits from such experiments exclude chameleon-photon couplings of $5\times10^{11}<\beta_\gamma<6.2\times10^{12}$ for effective chameleon masses less than 1 meV \cite{chou:030402}.  The vacuum chameleon masses covered by this type of experiment are model dependent, since the presence of residual gas and the nearby container walls change the effective chameleon mass inside the container.
%	D) Two sentence Quick Overview of how this will work

	An alternative technique is to trap chameleons inside a microwave cavity.
	Microwave cavities operate at lower energies than lasers, but have the advantage that the resonant nature of the cavity enhances the conversion probability between photons and chameleons.  As a result, microwave cavity experiments can potentially be more sensitive to $\beta_\gamma$.  We use the Axion Dark Matter Experiment (ADMX) to demonstrate this potential improvement for the first time.

%II. Chameleon Search with ADMX
%	A) ADMX design

	ADMX is a cavity search for dark matter axions \cite{PhysRevLett.51.1415}.  These axions are a consequence of the Peccei-Quinn solution to the strong CP problem \cite{PhysRevLett.38.1440,PhysRevLett.40.223,PhysRevLett.40.279}.  A full description of ADMX can be found in Ref.~\cite{Peng2000569}, and recent results from this axion search in Ref.~\cite{PhysRevLett.104.041301}.  In brief, ADMX consists of a 220 liter cylindrical copper-plated microwave cavity situated inside a 7 tesla magnet.  The cavity is held under vacuum and maintained at a temperature of 2 K produced by pumping on liquid helium.  Two copper rods are used to tune the resonant frequencies of the cavity.  When the $\mathrm{TM_{010}}$ resonant frequency of the cavity is tuned to correspond to the axion mass, the resonant mode will be excited by photons produced by the decay of dark matter axions.  The excitation of the resonant mode will be detected by an antenna probe inside the cavity and amplified by a microwave receiver.
	
	As with axions, chameleons can mix with photons in the microwave cavity.  Unlike axions, the chameleon mechanism traps the chameleon scalars inside the cavity along with the photons \cite{jaeckel:013004}.  
In this case the cavity will contain both electromagnetic resonances and chameleonic resonances, and the two will mix.  Consequently, the same technology ADMX uses to search for axion to photon conversion can be used to search for chameleon afterglow.

	 For the purposes of this analysis, it is assumed that the effective chameleon mass in the walls of the microwave cavity is much larger than the effective mass inside the cavity, yielding Dirichlet boundary conditions on the wave function.  Model dependent effects can modify this assumption, shrinking the effective cavity radius for chameleons, and changing their effective mass.  For a detailed analysis of chameleon behavior in cavities, see Ref.~\cite{brax:085010}.

	 Chameleon and photon mode mixing is maximized when the modes have the same frequency.  In the present ADMX cavity, this mixing should be most easily achieved between the $\mathrm{TE_{011}}$ electromagnetic cavity mode, which can be tuned between 850 and 950 MHz, and the lowest chameleon mode, which has a frequency that is the sum in quadrature of the effective chameleon mass ($m_\phi$) and wave number ($k_\phi$).  The position of the tuning rods inside the cavity can be moved to change the $\mathrm{TE_{011}}$ mode frequency and the lowest chameleon mode wave number by different amounts, and thus can be used to probe different chameleon masses.

	The procedure used to search for chameleons in ADMX was as follows (Fig.~\ref{fig:experiment_process}):

	\begin{enumerate}
	\item The $\mathrm{TE_{011}}$ electromagnetic mode was excited by driving the antenna with an external power source with a frequency swept over 20 kHz, roughly the width of the cavity resonance, for 10 minutes.  During this time period, if a chameleon mode were to overlap with the $\mathrm{TE_{011}}$ mode, some energy would be transferred to the chameleon mode.

	\item The external source was switched off and the first stage amplifier of the experiment was switched on.  In the duration of this switching period (100 ms), the conventional electromagnetic modes decayed.

	\item The power spectrum within 20 kHz of the \TEmode\ cavity resonance was recorded for 10 minutes.  If a chameleon mode had been excited in the previous step, its decay could be visible as an electromagnetic mode excitation.

	\item The tuning rods in the cavity were moved to change the frequency of the \TEmode\ mode, making it sensitive to a slightly different range of effective chameleon masses

	\end{enumerate}

\begin{figure}
\begin{center}
\includegraphics[height=8.5cm,angle=90]{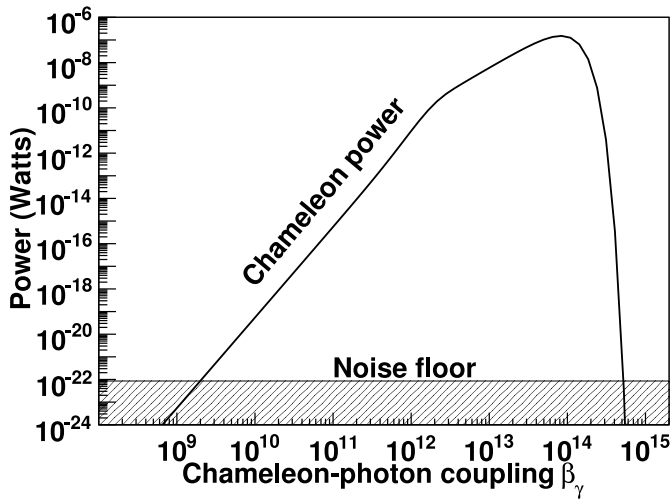} 
\caption{\label{fig:beta_comparison}Predicted excess power in the ADMX experiment from decaying chameleon modes immediately after turn on of the first stage amplifier as a function of chameleon-photon coupling strength $\beta_\gamma$.  Smaller $\beta_\gamma$ power is limited by coupling strength, larger $\beta_\gamma$ power is limited by decay time.  Noise floor corresponds to 3 K system temperature.}
\end{center}
\end{figure}

	Following the prescription used in Refs.~\cite{PhysRevLett.51.1415,PhysRevD.32.2988}, if $\beta_\gamma/\Mpl$ is sufficiently small, and the rate of chameleon loss to the cavity walls is negligible, the rate of mixing between the lowest chameleon mode mixing with the \TEmode\ mode would be

\begin{equation}
\Gamma=\frac{\beta_\gamma^2 f^2 B^2 Q k_{\mathrm{tr}}^2}{\Mpl^2 \omega^3},
\label{eqn:rate}
\end{equation}
where $\beta_\gamma$ and $\Mpl$ are as defined above, $f$ is a form factor (the overlap between the chameleon mode and the \TEmode\ mode which is roughly 0.43 in the case of ADMX), $B$ is the magnetic field strength, $Q$ is the cavity quality (around 10,000 for ADMX at this mode), $\omega$ is the driving frequency (around 900 MHz), and $k_{\mathrm{tr}}$ is the wave number of the chameleon mode transverse to the applied magnetic field.  The power detected in the cavity electromagnetic modes from chameleon decay would be

\begin{equation}
\Pout=\Pin \frac{\pi\Gamma}{2b} \left(1-e^{-\frac{\Gamma}{2} t_0}\right)^{2}e^{-\Gamma t},
\label{eqn:power}
\end{equation}
where $\Pin$ is the excitation power, $b$ is the bandwidth over which the driving frequency is swept (20 kHz in this experiment), $t_0$ is the duration for which the cavity has been excited, and $t$ is the time elapsed since the cavity excitation has ceased.  This is valid only when the sweep bandwidth is much larger than the chameleon resonance width, and the chameleon mode decay rate is smaller than the electromagnetic mode decay rate.

	The power excess would appear in the power spectrum as a peak at frequency $\omega=\sqrt{k_\phi^2+m_\phi^2}$ with a width $\Gamma$.  The power observed decreases exponentially with observation time as the chameleon mode decays with rate $\Gamma$.  The expected excess power in the ADMX experiment immediately after turning on the first stage amplifier as a function of coupling strength is shown in Fig.~\ref{fig:beta_comparison}.
	The signal-to-noise ratio of a chameleon signal, which determines its detectability, is given by the signal power (Eq.~\ref{eqn:power}) divided by the system noise temperature of the experiment.  The physical temperature of the ADMX cavity was 2 K at the time of data taking, and the SQUID (Superconducting Quantum Interference Device) amplifier\cite{Muck1998} had a noise temperature of 1 K, yielding a 3 K system noise temperature.
	
	The above discussion assumes that decay into photons is the dominant energy loss mechanism for excited chameleon modes.   Short range gravity experiments have limited the effective force between chameleons and matter to be weaker than the gravitational force, making energy loss to the walls negligible \cite{adelberger:131104}.
	Therefore, the vast majority of chameleons must eventually decay into photons.  As the wavelength of the chameleon mode is similar to that of the electromagnetic mode, both of which are much larger than the scale of any penetrations into the cavity, the bulk of the photons from chameleon decay are produced inside the cavity where they can be detected.  For this experiment, this translates to an assumption that the chameleon decay rate through means other than mixing with photons is less than $10^{-3}$ Hz, or an unloaded Q of greater than $10^{12}$, not far from that achievable in superconducting microwave cavities \cite{4325998}.

	There are two ways for a chameleon signal to be missed.  First, if the coupling is too weak, too little power is transferred from the electromagnetic mode to the chameleon mode and back to be detected.  Second, if the coupling is too strong, the chameleon mode can completely decay away in the time between the turn-off of the excitation and turn-on of the amplifier, and be indistinguishable from the the decay of the electromagnetic mode.

%	F) Results
\begin{figure}
\begin{center}
\includegraphics[width=6.5cm,angle=90]{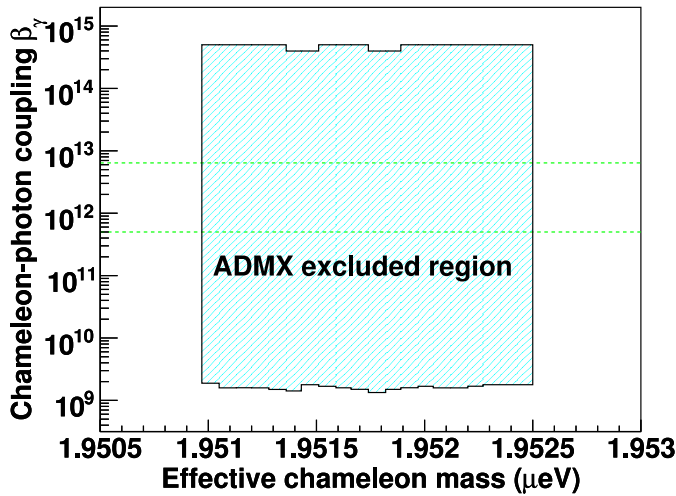}
\caption{\label{fig:admx_limits} Shaded region: 90\% confidence limit excluded parameters for scalar chameleons from ADMX search.  Dashed lines: upper and lower exclusion bounds from Ref.~\cite{chou:030402}.}
\end{center}
\end{figure}

	The procedure outlined here was followed with the ADMX experiment using a 25 dBm oscillator as the excitation source with a total integration time of $3.6\times10^4$ seconds.  The attenuation from the source to the cavity antenna was measured to be 28 dB, making the excitation power about 0.5 mW.  Given that no statistically significant power excess was observed, chameleon-photon couplings of $2\times10^{9}<\beta_\gamma<5\times10^{14}$ could be excluded at 90\% confidence over a mass range spanning 1.9510 $\mu$eV to 1.9525 $\mu$eV, as shown in Fig.~\ref{fig:admx_limits}.  
	As a reminder, the mass listed in the figure is the effective mass in the cavity, where the magnetic field was 7.1 T and the pressure was $10^{-6}$ torr.  The exact relation between this effective mass and true chameleon vacuum mass depends on one's choice of model.

	Compared to previous limits, the limit set by ADMX improves the lower bounds on chameleon-photon coupling by several orders of magnitude, but is valid only for a narrow range of effective masses.  The range of masses explored was limited by the time spent running; with a longer time, more chameleon masses could be explored at a rate of $10^{-3}$ $\mu$eV per day at the same sensitivity.

%IV. Conclusion
	In summary, we used ADMX to demonstrate the viability of microwave cavity searches for chameleon scalars.  Couplings of $2\times10^{9}<\beta_\gamma<5\times10^{14}$ were excluded for chameleons with an effective mass in the cavity between 1.9510 and 1.9525 \muev .  This technique is sensitive only to a narrow range of masses at each tuning setting, so it is most useful if a precise theoretical prediction can be made, or to confirm potential positive signals seen in other chameleon searches, such as those performed with lasers or short-range gravity experiments.

	ADMX will be upgraded soon from a system noise temperature of 3 K to an improved noise temperature of 200 mK by cooling the cavity to 100 mK, reducing the black body noise, and by lowering the temperature of the SQUID amplifier to 200 mK \cite{aphysj9672001}.  With a modest increase in excitation power, this would lead to an improvement on the lower bound of chameleon-photon coupling by an order of magnitude.  Much stronger couplings could be probed by a faster RF switching technique or lower magnetic field.  Even smaller chameleon-photon couplings can be probed by exciting the cavity for a longer time, but this impacts the speed over which masses can be scanned by a factor of 100 for every factor of ten improvement in chameleon-photon coupling sensitivity.  An accurate prediction of chameleon mass is still necessary to complete a search in a timely manner. 

	The ADMX collaboration gratefully acknowledges support by the U.S. Department of Energy, Office of High Energy Physics under contract numbers DE-AC52-07NA27344 (University of Washington), DE-FG02-96ER40956 (Lawrence Livermore National Laboratory), and DE-FG02-97ER41029 (University of Florida).
Additional support was provided by Lawrence Livermore National Laboratory under the LDRD program.  Development of the SQUID amplifier (JC) was supported by the 
Director, Office of Science, Office of Basic Energy Sciences, Materials 
Sciences and Engineering Division, of the U.S. Department of Energy 
under Contract No. DE-AC02-05CH11231.

\bibliographystyle{h-physrev}
\bibliography{admx_chameleons}

\end{document}